\documentclass[prl,amsmath,amssymb,twocolumn,superscriptaddress]{revtex4-2}
\usepackage{amsmath}
\usepackage{amssymb}
\usepackage{amstext}
\usepackage{amsfonts}
\usepackage{amsxtra}
\usepackage{bm}
\usepackage[usenames]{color}
\usepackage{grffile}
\usepackage{soul}
\usepackage{epstopdf}
\usepackage{xcolor}
\usepackage{graphicx}
\usepackage{marvosym}
\usepackage{wasysym}

{%
\setlength{\fboxsep}{0pt}%
\setlength{\fboxrule}{1pt}%
}%
\begin{document}

\title{Excitations of a binary dipolar supersolid}

\author{W. Kirkby}
 \affiliation{
     Universit\"{a}t Innsbruck, Fakult\"{a}t f\"{u}r Mathematik, Informatik und Physik, Institut f\"{u}r Experimentalphysik, 6020 Innsbruck, Austria
 }

\author{Au-Chen Lee}
  \affiliation{
	Dodd-Walls Centre for Photonic and Quantum Technologies, Dunedin 9054, New Zealand
}
 \affiliation{
     Department of Physics, University of Otago, Dunedin 9016, New Zealand
 }

\author{D. Baillie}
  \affiliation{
	Dodd-Walls Centre for Photonic and Quantum Technologies, Dunedin 9054, New Zealand
}
 \affiliation{
     Department of Physics, University of Otago, Dunedin 9016, New Zealand
 }

\author{T. Bland}
 \affiliation{
     Universit\"{a}t Innsbruck, Fakult\"{a}t f\"{u}r Mathematik, Informatik und Physik, Institut f\"{u}r Experimentalphysik, 6020 Innsbruck, Austria
 }
 
 \author{F. Ferlaino}
 \affiliation{
     Universit\"{a}t Innsbruck, Fakult\"{a}t f\"{u}r Mathematik, Informatik und Physik, Institut f\"{u}r Experimentalphysik, 6020 Innsbruck, Austria
 }
  \affiliation{
 	Institut f\"{u}r Quantenoptik und Quanteninformation, \"Osterreichische Akademie der Wissenschaften, Innsbruck 6020, Austria
 }
 
 \author{P. B. Blakie}
  \affiliation{
	Dodd-Walls Centre for Photonic and Quantum Technologies, Dunedin 9054, New Zealand
 }
 \affiliation{
     Department of Physics, University of Otago, Dunedin 9016, New Zealand
 }
 
 \author{R. N. Bisset}
 \affiliation{
     Universit\"{a}t Innsbruck, Fakult\"{a}t f\"{u}r Mathematik, Informatik und Physik, Institut f\"{u}r Experimentalphysik, 6020 Innsbruck, Austria
 }


\begin{abstract}

We predict a rich excitation spectrum of a binary dipolar supersolid in a linear crystal geometry, where the ground state consists of two partially immiscible components with alternating, interlocking domains.
We identify three Goldstone branches, each with first-sound, second-sound or spin-sound character.
In analogy with a diatomic crystal, the resulting lattice has a two-domain primitive basis and we find that the crystal (first-sound-like) branch is split into optical and acoustic phonons.
We also find a spin-Higgs branch that is associated with the supersolid modulation amplitude.

\end{abstract}

\date{\today}
\maketitle

The engineering of crystal phonons---or quantized sound waves---is an important challenge for ultracold gases \cite{Ostermann2016spontaneous,guo2021optical}, not least as simulators of solids with their central role in governing a material's thermodynamic and electrical properties \cite{chaikin1995principles,kittel2005introduction}.
Phonons had long eluded periodic optical potentials with neutral atoms, owing to the infinite lattice stiffness.
Major milestones have now been reached with the realization of supersolids, inherently possessing both the dissipationless flow of superfluids and the elastic crystalline structure of solids.
Despite their prediction over 50 years ago \cite{andreev1969quantum,chester1970speculations,leggett1970can,boninsegni2012colloquium}, supersolids were realized only recently using ultracold gases with dipolar interactions \cite{Tanzi2019,Bottcher2019,Chomaz2019,norcia2021two} and spin-orbit coupling \cite{li2017stripe,putra2020spatial}.
Supersolid phonons are of interest for a range of systems \cite{Watanabe2012,Saccani2012,Kunimi2012,Macri2013,Li2013,Roccuzzo2019,Geier2023,Blakie2023}, and experiments have begun probing them in dipolar condensates \cite{tanzi2019supersolid,guo2019low,natale2019excitation,Tanzi2021Evidence,Norcia2022Can,Chomaz2023dipolar}.


\begin{figure}[t]
	\centering
	\includegraphics[width=\columnwidth]{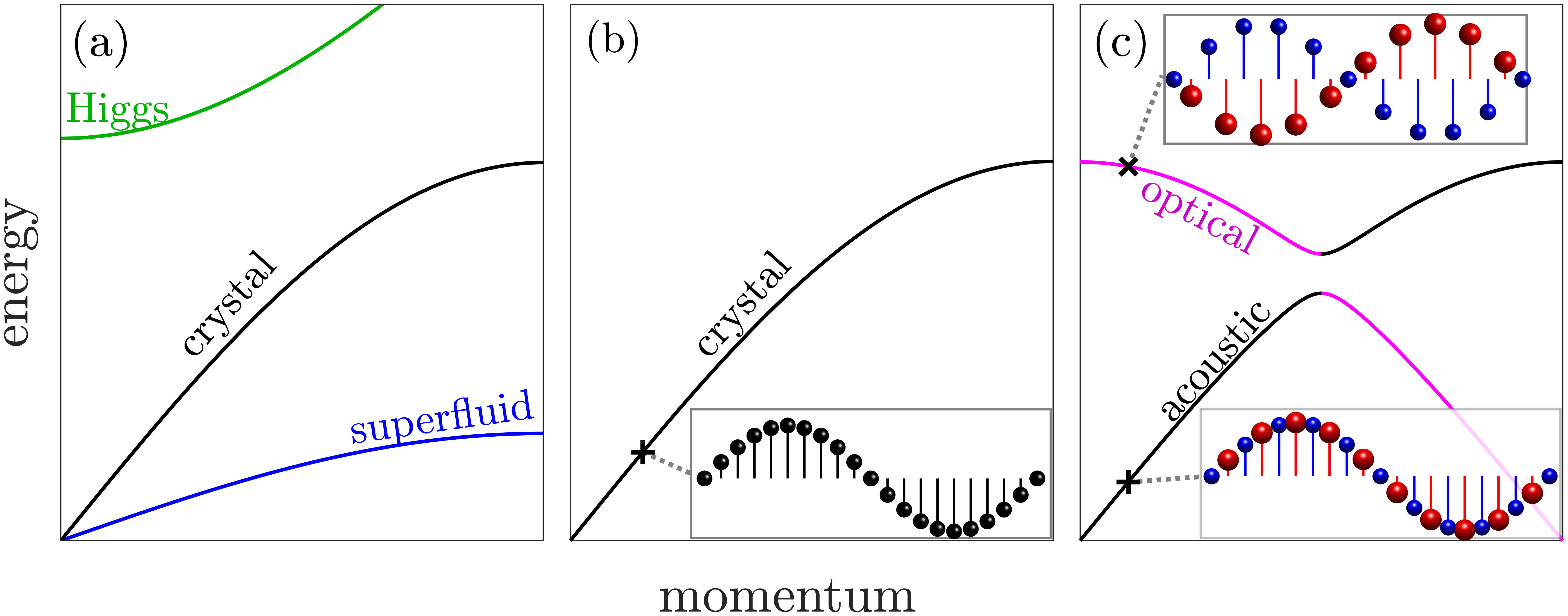}
	\caption{Schematic excitations for linear (a) single-component dipolar supersolid, and ordinary (b) monatomic and (c) diatomic crystals, with optical and acoustic phonons.}
	\label{fig:IntroFig}
\end{figure}


For linear supersolids, a crystal phonon branch [Fig.\ \ref{fig:IntroFig}(a)] appears similar to a regular solid [Fig.\ \ref{fig:IntroFig}(b)], emerging from a spontaneously broken continuous translational symmetry \footnote{Note that we have sketched atom displacements in the transverse direction for illustration purposes only. The present work deals exclusively with longitudinal excitations}.
A second Goldstone branch of superfluid phonons arises from breaking a global $U(1)$ gauge symmetry associated with the condensate order parameter \cite{Pitaevskii16}.
Higgs modes \cite{guo2019low,natale2019excitation,Hertkorn2019Fate}  
connected with a roton instability of the unmodulated phase \cite{ODell2003a,santos2003roton,Chomaz2018a,Petter2019,hertkorn2021density,schmidt2021roton} are also present.
There has been a significant undertaking to develop a general hydrodynamic model of supersolidity,
 e.g., see \cite{andreev1969quantum,Son2005,Yoo2010}, with compelling new advances \cite{Hofmann2021,Sindik2023sound}.
The periodic modulation of the density reduces the superfluid fraction \cite{leggett1970can} and---in analogy with finite-temperature superfluids---the lower phonon branch has a second-sound character, while the upper branch consists of first-sound-like modes \cite{Hofmann2021,Sindik2023sound}.
In dipolar supersolids, an intimate connection between the second-sound velocity and the superfluid fraction has been proposed as a practical means to measuring superfluidity \cite{Sindik2023sound}.

\begin{figure*}[t]
	\centering
	\includegraphics[width=\textwidth]{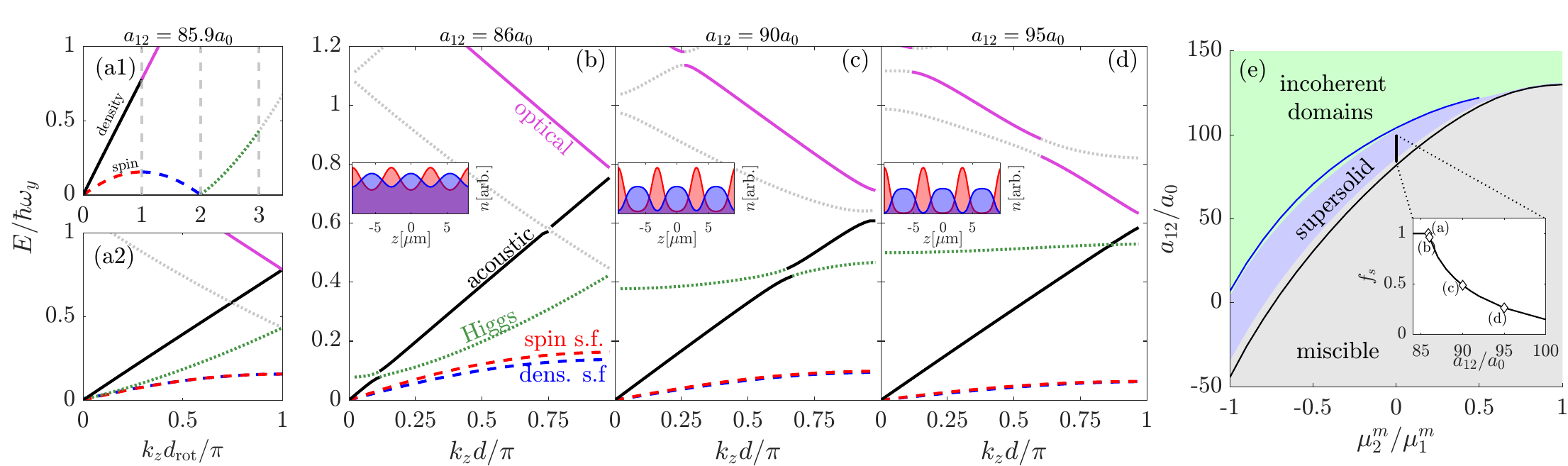}
	\caption{Binary supersolid dispersion.
	(a1) Excitation energies versus momentum for a uniform miscible state close to the transition.
	Sections of the spin (lower) and density (upper) branches have been color-coded, and these are folded into the first BZ in (a2) to demonstrate the origins of the supersolid branches that emerge in the next panels.
	(b)-(d) Supersolid excitations for increasing $a_{12}$, with insets showing integrated linear densities. (e) Phase diagram. Solid colors correspond to phases calculated using the variational method. For comparison, full 3D calculations are shown as solid lines, closely matching the corresponding uniform-supersolid (black) and supersolid-incoherent domain (blue) boundaries. Inset: superfluid fraction for a path across the transition (black vertical line) with the corresponding points for panels (a)-(d) indicated. }
	\label{fig:Evol}
\end{figure*}

Recent theory predicts the existence of binary dipolar supersolids, where two superfluids combine to form a periodically modulated state \cite{Bland2023Alternating,Li2023Long,politi2022interspecies,Scheiermann2023Catalyzation,kirkby2023spin,halder2023two}.
For a dipole imbalance between components, a special class forms partially immiscible, alternating domains, with the ground state stabilized by an interplay between the dipolar and contact interactions \cite{Bland2023Alternating,Li2023Long,kirkby2023spin}. 
Experimental advances with magnetic atoms suggest that binary supersolids may soon be realizable \cite{Trautmann2018,Ravensbergen2020,Durastante2020,politi2022interspecies,Schafer2023}, yet theoretical knowledge of the excitation spectrum is absent.
The two global gauge symmetries can spontaneously break independently, but intercomponent interactions mean that only a single translational symmetry (per spatial dimension) can spontaneously break, while the other is broken explicitly \cite{Watanabe2012}. Hence, for an $\mathcal{N}$-component supersolid in $D$ dimensions, one might expect $D+\mathcal{N}$ Goldstone modes.
%
The lattice of binary supersolids can be described by a multidomain primitive basis, and further understanding may be gleaned by analogy with nonsuperfluid crystals. 
In ordinary crystal lattices with two atoms per primitive cell, the phonon branch of a monatomic linear crystal [Fig.~\ref{fig:IntroFig}(b)] is replaced by a gapped spectrum in a diatomic lattice [Fig.~\ref{fig:IntroFig}(c)] \cite{ashcroft2022solid,kittel2005introduction}.
The size of the Brillouin zone (BZ) is thus halved and folded (magenta curves),
with the lower \emph{acoustic} (upper \emph{optical}) branch being characterized by in-phase (out-of-phase) motions between nearest neighbors [see insets].



In this Letter, we reveal the emergence and evolution of the excitation spectra of binary dipolar supersolids in an infinite tube.
From the uniform miscible phase, the spin branch---related to fluctuations driving phase
separation---develops rotonic excitations, and their instability is connected with the formation of the supersolid, having partially immiscible, alternating domains.
A rich excitation spectrum emerges with three Goldstone modes, where in addition to a second-sound-like branch there is a spin-sound branch, and the first-sound-like branch divides into optical and acoustic phonons. The modulation amplitude of the supersolid is associated with a spin-Higgs branch.

We consider a pair of distinguishable BECs with wavefunctions $\Psi_i$ ($i=1,2$), described by two coupled Gross-Pitaevskii equations (GPEs) \cite{SM}.
The atoms are trapped in an infinite tube potential oriented along the $z$ axis, $V_i(\textbf{r})=\frac{1}{2}m_i(\omega_x^2x^2+\omega_y^2y^2)$.
To remain firmly in the linear crystal regime we select trapping frequencies $\{\omega_x,\omega_y\}/2\pi=\{300,100\}$\,Hz \footnote{Reference \cite{Bland2023Alternating} showed that even in nearly circular traps the effects of magnetostriction can cause nontrivial density modulations along the trapped direction perpendicular to the dipole polarization axes.}. 
To highlight the novel features of a dipole-imbalanced mixture, we take balanced intraspecies scattering lengths $a_{ii}=130a_0$, average linear densities $\bar{n}_i=200\mu\mathrm{m}^{-1}$ and the mass of ${}^{164}\mathrm{Dy}$, $m_i=164\mathrm{u}$. For the dipoles, which are polarized along $y$, we take $\mu_1^m=9.93\mu_\mathrm{B}$ and unless otherwise stated, we consider the second component as nondipolar $\mu_2^m=0$.  
We generalize a 3D variational theory developed in Ref.\ \cite{Blakie2020variational} to two components. Decomposing the wavefunctions as $\Psi_i(\textbf{r})= \psi_i(z)\phi_i(x,y)$, we then assume radial degrees of freedom are described by
$\phi_i(x,y)=\frac{1}{\ell_{i}\sqrt{\pi}}\exp\left[-(\eta_{i} x^2+y^2/\eta_{i})/(2\ell_{i}^2)\right]$,
where $\eta_i$ and $\ell_i$ are variational parameters corresponding to the aspect ratio and average width, respectively, which are determined by minimizing the energy.
The ground state is calculated by solving the two-component GPEs in imaginary time. By making use of Fourier copies along $z$, only a single primitive unit cell needs to be simulated to reach the ground state in the thermodynamic limit, while the crystal lattice spacing $d$ is varied until the energy is minimized. The excitation spectrum is calculated using a Bogoliubov-de Gennes (BdG) formalism \cite{SM}. 

Longitudinal excitation spectra within the first BZ are shown in Fig.~\ref{fig:Evol} for various interspecies coupling constants $a_{12}$, ranging from near the supersolid transition point [Fig.~\ref{fig:Evol} (a)]---where the density is still uniform---to deep inside the supersolid regime [Fig.~\ref{fig:Evol} (d)].
Mode types are labeled in panel (b), showing a Goldstone acoustic phonon branch and two Goldstone superfluid (s.f.) branches, where the latter will later be distinguished as being either spin or density dominated. 
Spin Higgs and optical branches are also labeled, while higher superfluid modes that are not the central focus of this work are plotted as gray dotted lines.
As $a_{12}$ is increased, the reduction of intersite superfluidity is apparent in the linear density plots (insets), where the dipolar (nondipolar) component is shown in red (blue). The increasingly isolated domains act to flatten the bands of the superfluid modes [dotted/dashed lines in Fig.~\ref{fig:Evol} (d)], leaving the nontrivial optical and acoustic crystal branches to more closely resemble a solid without superfluidity [Fig.~\ref{fig:IntroFig} (c)].

To understand the origin of each branch, Fig.~\ref{fig:Evol} (a1) shows the excitation spectrum in the unmodulated miscible phase very close to the transition. The density branch remains hard, while the spin branch has developed a roton minimum that softens (lowers) and becomes unstable as the transition is crossed. Defining a fictitious BZ boundary for a lattice vector corresponding to the roton wavelength, $d_{\mathrm{rot}}$, we then color parts of both branches as they cross through each BZ, with the BZ `edges' shown as vertical dashed lines. Figure \ref{fig:Evol} (a2) shows the same data, but with the spectrum folded into the reduced zone scheme. From this mapping, it becomes clear that crystal modes emerge from the density branch, with the lower (upper) part becoming the acoustic (optical) branch. In contrast, superfluid modes emerge from the spin branch, with the spin Higgs mode connecting with the unstable spin rotons at the transition. 

Situating our system within the context of a larger available parameter space, we show in Fig.~\ref{fig:Evol} (e) a phase diagram for a range of dipolar mixtures (fixed $\mu_1^m=9.93\mu_{\rm B}$ but varying $\mu_2^m$). We classify phases using the upper bound on the superfluid fraction as outlined by Leggett, $f_{s}^{(i)}=\frac{d^2}{N_i}\left[\int_{\mathrm{uc}}\mathrm{d}z\left(\int\mathrm{d}x\mathrm{d}y|\Psi_i|^2\right)^{-1}\right]^{-1}$, where $N_i=\bar{n}_id=\int_{\mathrm{uc}}\mathrm{d}z|\psi_i|^2$ is the particle number of component $i$ within a unit cell (uc), such that the reduction in total moment of inertia of the composite system is related to the total superfluid fraction, $f_s=(N_1f_s^{(1)}+N_2f_s^{(2)})/(N_1+N_2)$.
In the miscible phase (gray), there is no density modulation in the region immediately below the supersolid phase and $f_{s}=1$ \footnote{Note that a miscible supersolid phase is expected in the bottom right corner of the phase diagram \cite{Scheiermann2023Catalyzation,Bland2023Alternating} (not shown), which is not the focus of the present work.}. 
Increasing $a_{12}$ leads to a transition to the supersolid ($0.1<f_s<1$) phase and subsequently to the incoherent domain ($f_s\leq 0.1$) regime.
The supersolid phase can be seen to extend over a broad range of dipole combinations, 
and we have checked (not shown) that our excitation findings are qualitatively representative throughout this region.
We also perform full 3D ground state calculations (solid curves) to corroborate our predictions for the uniform-to-supersolid and supersolid-to-incoherent domain phase boundaries, and find good agreement throughout the phase diagram.
The inset shows how $f_s$ changes as the transition is crossed for the dipolar-nondipolar system, with the locations of the preceding panels of Fig.\ \ref{fig:Evol} also indicated.




\begin{figure}[t]
	\centering
	\includegraphics[width=\columnwidth]{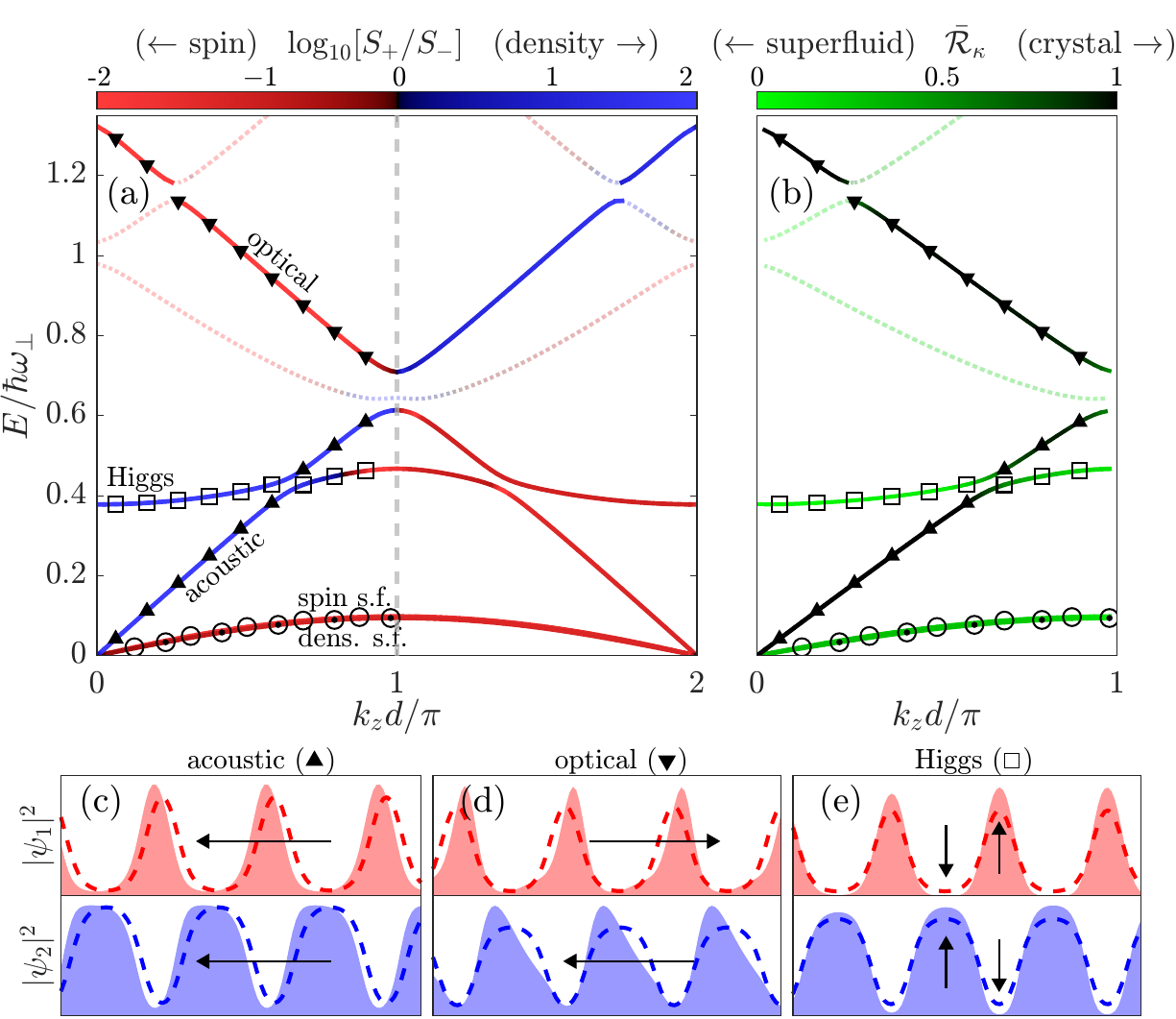} 
	\caption{Branch identification for dipolar-nondipolar mixture with $a_{12}=90a_0$. (a) Excitation branches are shown and colored based on the relative strength of the dynamic structure factors. Symbols to guide the eye indicate acoustic ($\blacktriangle$) and optical ($\blacktriangledown$) crystal branches, a spin Higgs ($\square$) branch, as well as spin- ($\ocircle$) and density- ($\odot$) superfluid branches. (b) Superfluid versus crystal character (see main text). (c)-(e) Examples of long-wavelength modes, where one can see the ground state (dashed) and the same state with the corresponding BdG excitation added (shaded) (see supplementary videos \cite{SM}). }
	\label{fig:BranchID}
\end{figure}


The lower panels of Fig.\ \ref{fig:BranchID} correspond to exemplary BdG modes of the binary supersolid system. The ground state over a few lattice sites is shown with dashed lines, and the same states including the corresponding excitations are shown with shading. The BdG modes are included with a sufficiently large amplitude to emphasize the motion. The acoustic and optical modes are dominated by spatial oscillations of the domains, in-phase and out-of-phase, respectively, while for the spin Higgs mode, the domains remain stationary but the spin-density amplitude ($||\psi_2|^2-|\psi_1|^2|$) changes.

For a quantitative understanding, we consider two measures of excitation branch identification. In Fig.~\ref{fig:BranchID} (a) we plot the dynamic structure factor \cite{abad2013study,symes2014static},
\begin{equation}
    S_{\pm}(k_z,\omega) =\frac{1}{N} \sum_{\kappa }\left|\delta \tilde{n}^{\pm}_{\kappa }(k_z)\right|^2\delta(\omega-\omega_\kappa ) \, , \label{e:Spm}
\end{equation}
where $\delta \tilde{n}^{\pm}_{\kappa }(k_z)=\int_0^L\mathrm{d}z\;\mathrm{e}^{-\mathrm{i}k_zz}[\delta n_{\kappa ,1}(z)\pm\delta n_{\kappa ,2}(z)]$, the energy of excitation $\kappa $ is $\omega_\kappa =E_\kappa/\hbar$ and the density fluctuations are $\delta n_{\kappa,i}(z)=\psi_i(z)[u_{\kappa,i}(z)-v_{\kappa,i}(z)]$, with the modes $u_{\kappa,i}$ and $v_{\kappa,i}$ being the BdG amplitudes \cite{SM}. In determining branch character, we simulate a larger system of size $L=N_{\mathrm{uc}}d$ for $N_{\mathrm{uc}}=96$, and total particle number $N=N_{\mathrm{uc}}(N_1+N_2)$. The spectrum is colored by the relative strength of the density $(+)$ or spin $(-)$ dynamic structure factor along each branch, with blue representing a stronger signal in $S_+$, and red for $S_-$. The density structure factor dominates when the components oscillate in-phase, while the spin structure factor is stronger for out-of-phase fluctuations. 
In Fig.~\ref{fig:BranchID} (b) we plot $\bar{\mathcal{R}}_\kappa=(\mathcal{R}_{\kappa,1}+\mathcal{R}_{\kappa,2})/2$ as a measure of the crystal versus superfluid character of each quasi-particle excitation labeled by quantum number $\kappa$, where \cite{SM}\footnote{See Ref.~\cite{natale2019excitation} for a related calculation for a single-component system.}
\begin{equation}
	\mathcal{R}_{\kappa,i}=\frac{\frac{1}{N_{\mathrm{uc}}}\sum_{p}^{N_{\mathrm{uc}}}|\mathrm{v}_{\kappa,i}(z_p)|}{\frac{1}{L}\int_0^L\mathrm{d}z\;|\mathrm{v}_{\kappa,i}(z)|}\;, \label{Eq:Rvi}
\end{equation}
with $L=N_{\mathrm{uc}}d$ being the total length of our simulated system and $z_p$ is the location of density peak $p$. The induced superfluid velocity, $\mathrm{v}_{\kappa,i}=
\frac{\hbar}{m_i}\mathrm{d}\delta \varphi_{\kappa,i}/\mathrm{d}z$, can be calculated from the local phase fluctuation $	\delta\varphi_{\kappa,i}(z)=[u_{\kappa,i}(z)+v_{\kappa,i}(z)]/\psi_i(z)$. The quantity $\bar{\mathcal{R}}_\kappa$ compares the average motion caused by the excitation around the density peaks of each component relative to the interstitial regions, and is therefore sensitive to its crystal ($\mathcal{\bar R}_{\kappa}\gtrsim 1$) or superfluid ($\mathcal{\bar R}_{\kappa}\ll 1$) character. 

Avoided crossings between the branches arise from couplings between the modes  (cf. Refs.~\cite{petter2020high,Blakie2023}), resulting in hybridization of mode character in these regions. 
Nevertheless, both the optical and acoustic branches can clearly be identified by their strong crystal character [black curves in Fig.~\ref{fig:BranchID} (b)], and they can be distinguished from one another by the relative strength of the density or spin structure factors [Fig.~\ref{fig:BranchID} (a)]. The strong density character of the acoustic branch implies that the crystal excitations of both components occur in the same direction, i.e., neighboring domains move in phase [Fig.\ \ref{fig:BranchID} (c)], while for the optical branch (spin dominated) the components oppose one another [Fig.\ \ref{fig:BranchID} (d)].





\begin{figure}[t]
	\centering
	\includegraphics[width=0.8\columnwidth]{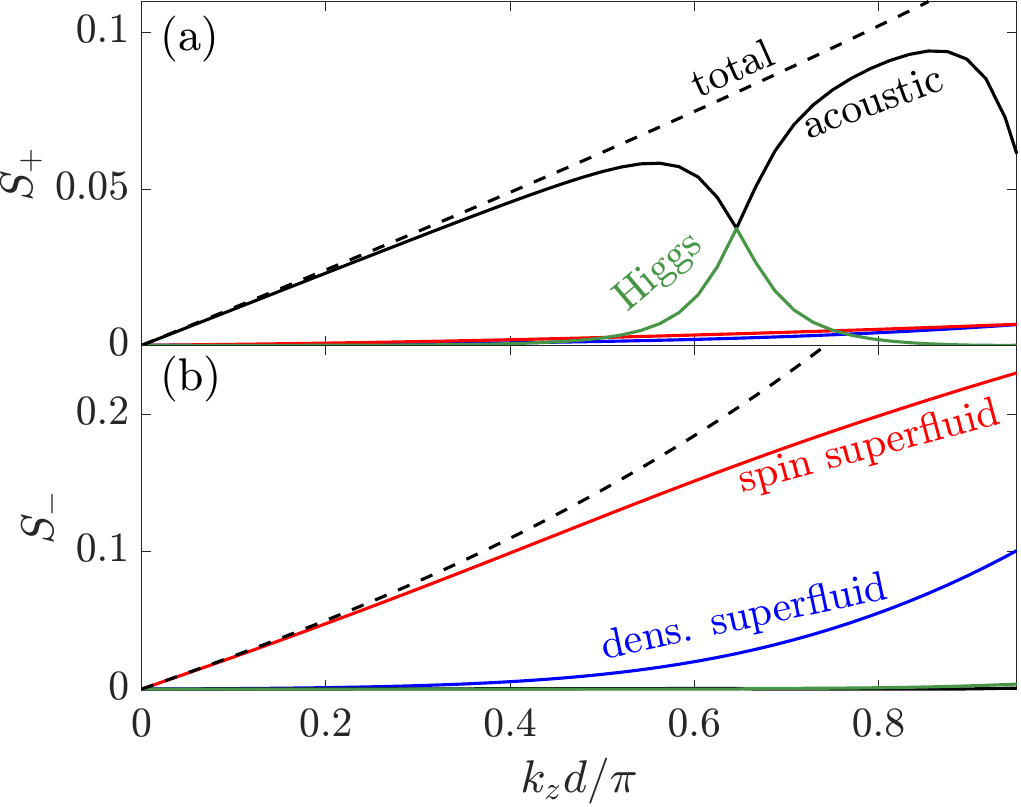}
	\caption{Goldstone mode contributions to the static structure factor for a dipolar-nondipolar mixture at $a_{12}=90a_0$. (a) Density and (b) spin static structure factors shown as black-dashed lines, with the contributions from relevant branches (solid) color-coded for comparison with Fig.~\ref{fig:Evol} (c).}
	\label{fig:Static}
\end{figure}


The two-component nature of our system (i.e.~two domains per unit cell) opens new possibilities for addressing excitations. 
Notably, if an excitation predominantly couples to the spin structure factor for momentum transfer in the first BZ, in the second BZ it may predominantly couple to the density structure factor.
For example, Fig.\ \ref{fig:BranchID} (a) shows the spin Higgs excitations to have a strong signal in $S_+$ within the first BZ, yet these same modes are dominated by $S_-$ in the second BZ, where they modify the spin modulation amplitude at the roton wavelength. 
The situation is reversed for the optical branch, which emerges from the density branch in the unmodulated regime (Fig.~\ref{fig:Evol}), but has strong spin character in the first BZ [Fig.~\ref{fig:BranchID} (a)]. 
The spin-density flipping can be understood by considering balanced systems with equal populations and intracomponent interactions, e.g., with $\mu_1^m=-\mu_2^m$. In this limit, excitations will have strictly density (or spin) character within the first BZ, but this character will change to spin (or density) in the second BZ. More generally, there is a flipping of density-spin character for every reciprocal lattice vector \cite{SM}. For imbalanced systems the spin and density modes couple and their character can hybridize. 
An interesting feature of our system is that bands higher than the optical branch may be used to probe the internal structure of individual lattice sites, whereas such bands are absent in idealized solids with point-like particles.

To differentiate the characteristics of the three Goldstone branches, we turn to the density and spin static structure factors defined as $S_{\pm}(k_z)\equiv \int\mathrm{d}\omega\; S_{\pm}(k_z,\omega)$.
Figure \ref{fig:Static} (a) shows the density structure factor (dashed line), along with the contributions from each the three Goldstone branches, and the Higgs branch, colored to match the corresponding energy branches in Fig.~\ref{fig:Evol} (c).
While the density static structure factor is dominated by the acoustic branch (black solid line) at low and high momenta, there is significant mixing close to the avoided crossing between the Higgs and acoustic branches.
In the spin static structure factor [Fig.~\ref{fig:Static} (b)], there is a strong distinction in the contributions from the two superfluid Goldstone branches, allowing us to distinguish the ``spin" superfluid branch from the ``density" superfluid branch. Interestingly, both superfluid Goldstone branches emerge from the spin branch in the unmodulated regime, with the folding of the density branch from the 2nd BZ to the 1st BZ [Fig.~\ref{fig:Evol} (a)] responsible for the shift to density character.
Further, we have checked that the density (spin) branches create superfluid flows in the two components that align with (oppose) one another \cite{SM}. 
Closer to the phase transition (not shown), the distinction becomes less clear and both low superfluid branches contribute with comparable weight to the structure factors.

In summary, we have performed the first study of the excitation spectrum of a binary dipolar supersolid, focussing on an elongated geometry.
Three Goldstone branches emerge due to the spontaneous breaking of one translational symmetry and two gauge symmetries, confirming the anticipated $D+\mathcal{N}$ Goldstone modes of multicomponent supersolids.
In addition to Goldstone modes with first- and second-sound character, we identify a Goldstone branch with spin-sound character.
Further, in analogy with ordinary solids, the spectrum exhibits both optical and acoustic phonon branches, and in addition, a spin Higgs branch emerging from the spin roton.
It is worth noting that spin-orbit coupled supersolids exhibit only two Goldstone branches, as coupling between spin and translational degrees of freedom reduces the number of spontaneously broken symmetries, similar to the single-component supersolid \cite{Geier2023,Liao2018,Martone2021}.

We expect our findings to be attainable with current experimental capabilities, thanks to their generality across a wide parameter space of imbalanced mixtures, including heteronuclear combinations of magnetic atoms \cite{Trautmann2018,Ravensbergen2020,Durastante2020,politi2022interspecies,Schafer2023} and homonuclear mixtures with various spin projections \cite{Chalopin2020}. 
Building on successful strategies for single-component supersolids \cite{petter2020high,Petter2019,guo2019low,hertkorn2021density,tanzi2019supersolid}, the structure factor may be accessed via density and spin Bragg spectroscopy techniques \cite{Hoinka2012}, or via component-sensitive {\it in-situ} imaging.
Since immiscible binary supersolids do not require quantum fluctuations for stabilization, in contrast to other dipolar supersolids, we expect lower densities and crystals with significantly more lattice sites.
Ideal geometries include strongly prolate trapping or toroidal traps \cite{Gupta2005,Olson2007,Ryu2007,Henderson2009,Tengstrand2021,Tengstrand2023}.
Our work opens intriguing perspectives for future work to develop a general hydrodynamic model and an effective Lagrangian for $\mathcal{N}$-component supersolids \cite{andreev1969quantum,Son2005,Yoo2010,Hofmann2021,Sindik2023sound}.

{\it Acknowledgements:---}
R.~B., T.~B., F.~F. and W.~K.~acknowledge financial support by the ESQ Discovery programme (Erwin Schr\"odinger Center for Quantum Science \& Technology), hosted by the Austrian Academy of Sciences (\"OAW). D.~B., P.~B.~B. and A.~C.~L.~ acknowledge support from the Marsden Fund of the Royal
Society of New Zealand. R.~B. also acknowledges the Austrian Science Fund (FWF): P 36850-N. F.~F. and W.~K. acknowledge support from the European Research Council through the Advanced Grant DyMETEr (No. 101054500). F.~F. acknowledges support the DFG/FWF via Dipolare E2 (No. I4317-N36) and, with T.~B., a joint project grant from the FWF (No. I4426).

\bibliography{DipolarRefs}

\section{Supplementary Materials}

\renewcommand{\theequation}{S\arabic{equation}}
\setcounter{equation}{0} 
\renewcommand{\thefigure}{S\arabic{figure}}
\setcounter{figure}{0} 

\subsection{GPE and Variational theory}
The two-component dipolar GPE is given by,
\begin{align}
	\mathrm{i}\hbar\frac{\partial \Psi_i(\textbf{r})}{\partial t}=&\Biggl[-\frac{\hbar^2\nabla^2}{2m_i}+V_i(\textbf{r})+\sum_{j}g_{ij}n_{j}(\textbf{r})\nonumber\\&+\sum_{j}\int\mathrm{d}\textbf{r}'U_{ij}(\textbf{r}-\textbf{r}')n_{j}(\textbf{r}')\Biggr]\Psi_i(\textbf{r})\;,\label{eq:GPE}
\end{align}
with atomic mass $m_i$, density $n_i(\textbf{r})=|\Psi_i(\textbf{r})|^2$, contact interaction strength $g_{ij}=2\pi \hbar^2a_{ij}(m_i+m_j)/(m_im_j)$, and trapping potential $V_i(\textbf{r})=\frac{1}{2}m_i(\omega_x^2x^2+\omega_y^2y^2)$. The dipole-dipole interaction (DDI) between particles of magnetic moment $\mu^m_i$ and $\mu^m_j$ is given by
$U_{ij}(\textbf{r})=\mu_0\mu_i^m\mu_j^m(1-3\cos^2\theta)/4\pi|\textbf{r}|^3$, where
$\theta$ is the angle between the vector $\textbf{r}$ joining two dipoles and the polarization direction, which we take to be the $y$ axis, and $\mu_0$ is the vacuum permeability. 

As described in the main text, we consider a reduced 3D variational theory by assuming the wavefunction takes the form, $\Psi_i(\textbf{r})= \psi_i(z)\varphi_i(x,y)$, with $\varphi_i(x,y)=\frac{1}{\ell_{i}\sqrt{\pi}}\exp\left[-(\eta_{i} x^2+y^2/\eta_{i})/(2\ell_{i}^2)\right]$. The two variational parameters can also be written in terms of the $1/\mathrm{e}$ half-widths $l^{x,y}$ of the density $|\varphi_i(x,y)|^2$ along each axis via $\ell_i=\sqrt{l_i^xl_i^y}$ and $\eta_i=l_i^y/l_i^x$. We then multiply the GPE by $\varphi_i(x,y)$ and integrate over the azimuthal directions. 

The integrated quasi-1D GPE is
\begin{align}
	\mathrm{i}\hbar\frac{\partial \psi_i(z,t)}{\partial t}=&\;\Biggl[-\frac{\hbar^2}{2m_i}\frac{\partial^2}{\partial z^2}+\mathcal{E}_{\perp}^i+\sum_{j}\frac{g_{ij}}{2 \pi \ell_{ij}^2}|\psi_{j}(z,t)|^2\nonumber\\&+\sum_{j}\int\mathrm{d}z'U_{ij}^{1\mathrm{D}}(z-z')|\psi_{j}(z',t)|^2\Biggr]\psi_i(z,t) ,
\end{align}
where we have defined inter-component effective variational widths such that
\begin{align}
	4\ell_{ij}^4=&\;(\ell_i^2\eta_i+\ell_j^2\eta_j)(\ell_i^2/\eta_i+\ell_j^2/\eta_j)\\
	\eta_{ij}^2=&\;\frac{\ell_i^2\eta_i+\ell_j^2\eta_j}{\ell_i^2/\eta_i+\ell_j^2/\eta_j} ~,
\end{align}
from which it follows naturally that $\ell_{ii}=\ell_i$ and $\eta_{ii}=\eta_i$. The energy per particle in component $i$ due to the trapping potential is given by
\begin{equation}
	\mathcal{E}_{\perp}^i=\frac{\hbar^2}{4m_i\ell_i^2}\left(\eta_i+\frac{1}{\eta_i}\right)+\frac{1}{4}m_i\ell_i^2\left(\frac{\omega_x^2}{\eta_i}+\eta_i\omega_y^2\right)\;.
\end{equation}
For the integrated quasi-1D DDI, we make use of the expression developed in Ref.\  \cite{blakie2020supersolidity} generalized to multiple variational widths
\begin{align}
	\mathcal{F}[U_{ij}^{1\text{D}}(z)]=&\;\frac{\mu_0\mu_i^m\mu_j^m}{6\pi\ell_{ij}^2}\biggl[\frac{2-\eta_{ij}-3\cos^2\alpha}{1+\eta_{ij}}\nonumber\\&+A_\alpha Q_{ij}^2\mathrm{e}^{Q_{ij}^2}\mathrm{Ei}\left(-Q_{ij}^2\right)\biggr] ~,
\end{align}
where $\mathcal{F}$ denotes the Fourier transform, $\mathrm{Ei}(x)$ is the exponential integral, $A_\alpha=3\left(\frac{1-\cos^2\alpha}{1+\eta_{ij}}-\cos^2\alpha\right)$, $Q_{ij}(\alpha=\pi/2)=\frac{1}{\sqrt{2}}\eta_{ij}^{\frac{1}{4}}k\ell_{ij}$, and $Q_{ij}(\alpha=0)=\frac{1}{\sqrt{2}}\left(\frac{2\eta_{ij}}{1+\eta_{ij}^2}\right)^{\frac{2}{5}}k\ell_{ij}$ for dipole polarization angle $\alpha$.

\subsection{BdG equations}

Starting from the integrated quasi-1D GPE, we consider perturbations to the real ground state $\psi_i$ of the form
\begin{equation}
    \tilde\psi_i(z,t)=\left\{\psi_i(z)+\lambda\left[u_i(z)\mathrm{e}^{-\mathrm{i}\omega t}-v_i^*(z)\mathrm{e}^{\mathrm{i}\omega^* t}\right]\right\}\mathrm{e}^{-\mathrm{i}\mu_i t/\hbar} , \label{e:GPEz}
\end{equation}
where $\lambda$ is some small real perturbative parameter. Defining the linear operators $L_i$ and $X_{ij}[\;\cdot\;]$, acting on a function $w$,
\begin{align}
	L_i w_i\equiv&\;-\frac{\hbar^2}{2m_i}\frac{\partial^2}{\partial z^2}w_i(z)+\mathcal{E}_{\perp}^iw_i(z)\nonumber\\
	&+\sum_{j}\int\mathrm{d}z'\;W_{ij}(z-z')|\psi_{j}(z')|^2w_i(z) \\
	X_{ij}[w_{j}]=&\;\psi_{i}(z)\int\mathrm{d}z'\;\psi_{j}(z')W_{ij}(z-z')w_{j}(z') \ ,
\end{align}
with interaction matrix
\begin{equation}
	W_{ij}(z-z')=\frac{g_{ij}}{2\pi\ell_{ij}^2}\delta(z-z')+U_{ij}^{1\text{D}}(z-z')\label{eq:InteractionMatrix}\;,
\end{equation}
gives the Bogoliubov-de Gennes equations
\begin{align}
	(\hbar \omega +\mu_i)u_i(z)=&\;L_i u_i(z)+\sum_j\left(X_{ij}[u_j]-X_{ij}[v_j]\right)\\
	(\hbar \omega -\mu_i)v_i(z)=&\;-L_i v_i(z)+\sum_j\left(X_{ij}[u_j]-X_{ij}[v_j]\right) ~.
\end{align}


\subsection{Superfluid versus crystal character}
\label{appdx:Character}

The local superfluid velocity  for a condensate written in the form $\Psi=\sqrt{n}\mathrm{e}^{\mathrm{i}\varphi}$ is given by \cite{Pitaevskii16}
\begin{equation}
	\mathrm{v}=\frac{\hbar}{m}\nabla \varphi\;.
\end{equation}
Assuming the phase is uniform across the supersolid ground state, the relevant contribution to the superfluid velocity for a state with a BdG excitation labeled by $\kappa$ is,
\begin{equation}
	\delta\varphi_{\kappa,i}(z)=[u_{\kappa,i}(z)+v_{\kappa,i}(z)]/\psi_i(z)\;.
\end{equation}
The gradient of $\delta \varphi_{\kappa,i}$ is proportional to the superfluid velocity via $\mathrm{v}_{\kappa,i}=
\frac{\hbar}{m_i}\mathrm{d}\delta \varphi_{\kappa,i}/\mathrm{d}z$.

We then use this to determine the mode character through the ratio
\begin{equation}
\mathcal{R}_{\kappa,i}=\frac{\frac{1}{N_{\mathrm{uc}}}\sum_{p}^{N_{\mathrm{uc}}}|\mathrm{v}_{\kappa,i}(z_p)|}{\frac{1}{L}\int_0^L\mathrm{d}z\;|\mathrm{v}_{\kappa,i}(z)|}\;, \label{Eq:Rvi_SM}
\end{equation}
where $L=N_{\mathrm{uc}}d$ is the total length of our simulated system, and $z_p$ is the location of the supersolid peak $p$. The numerator is an average over the speed contributions at the density peaks of every domain, indexed by $p$, where we take $N_{\mathrm{uc}}=96$. The ratio $\mathcal{R}_{\kappa,i}$ can thus be interpreted as a measure of the crystal or superfluid nature of each excitation by comparing the average motion of the crystal sites (numerator) to the average superfluid motion across the entire system (denominator).
Equation (\ref{Eq:Rvi_SM}) can be understood by realizing that a large wavefunction phase gradient at a density peak is associated with motion of the domain itself, contributing to a predominantly crystal character when $\mathcal{R}_{\kappa,i}\gtrsim 1$, whereas superfluid excitations are instead associated with fast superfluid currents within the low density regions between peaks, giving $\mathcal{R}_{\kappa,i}\ll 1$.
In Fig.~3 (b) of the main text, 
the coloring is determined by the average for the two components $\bar{\mathcal{R}}_\kappa=(\mathcal{R}_{\kappa,1}+\mathcal{R}_{\kappa,2})/2$. Using a two-component generalization of the measure $\mathcal{C}$ found in Ref. \cite{natale2019excitation} gives quantitatively similar results.

\begin{figure}
	\centering
	\includegraphics[width=0.8\columnwidth]{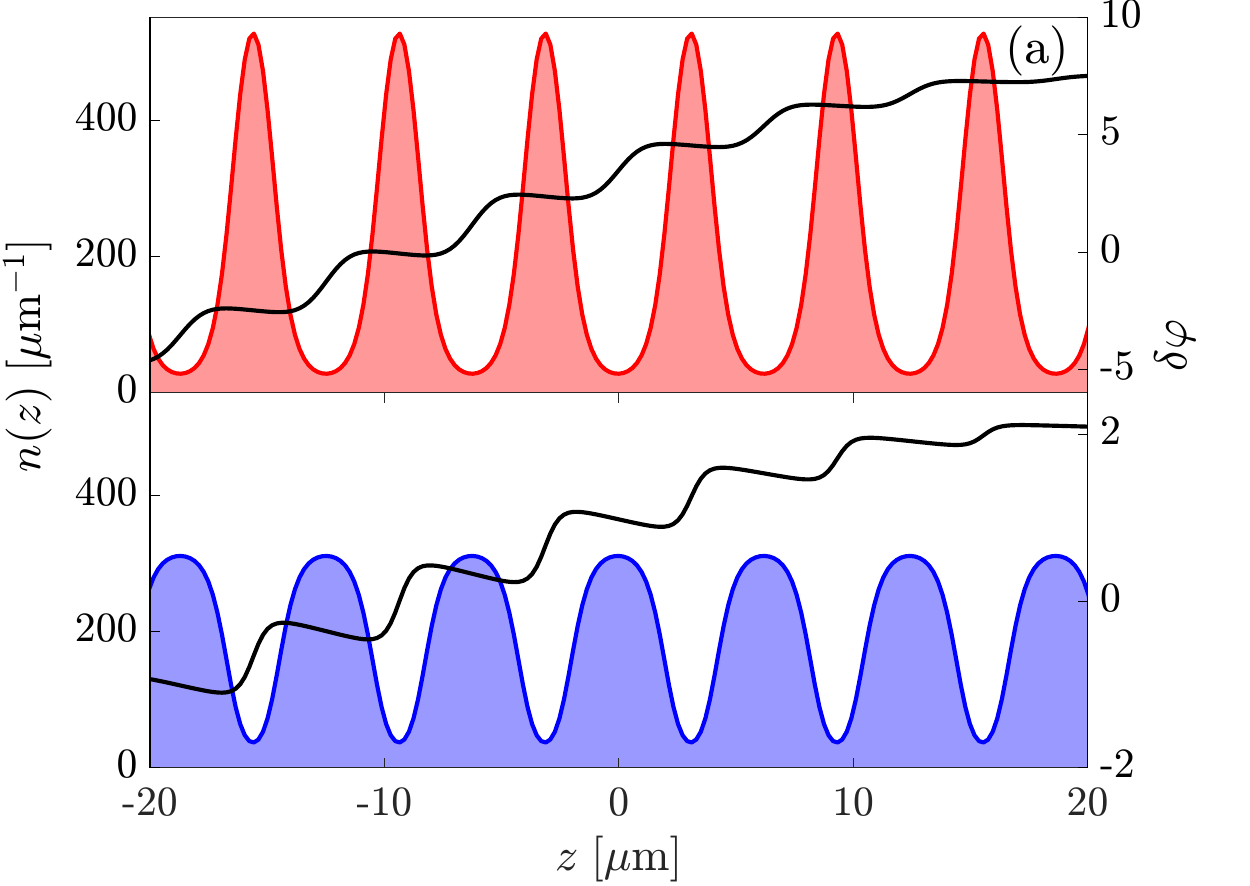}
	\includegraphics[width=0.8\columnwidth]{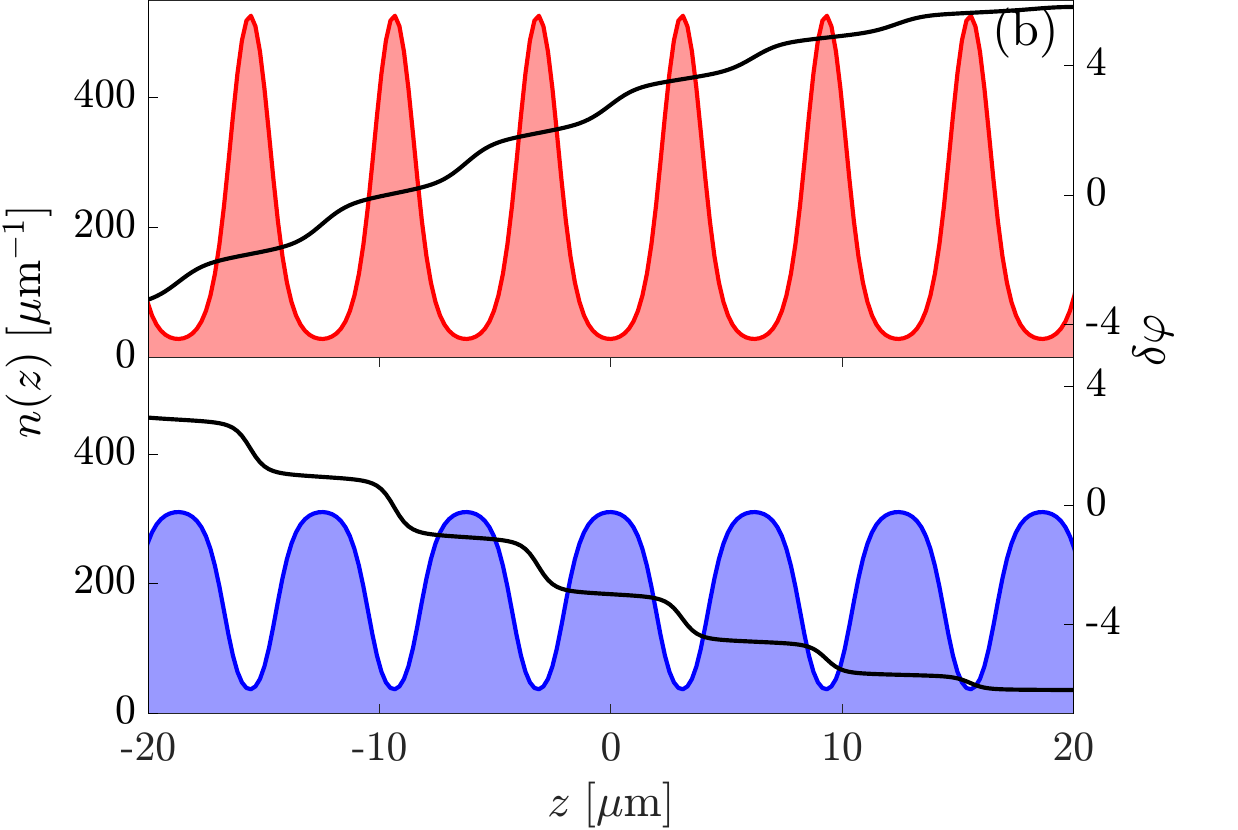}	
	\caption{Phase variation of the superfluid Goldstone branches at $k_zd/\pi\approx 0.10$. The ground state is shown in solid colors for the dipolar component in red, and the nondipolar component in blue. (a) Density superfluid mode and (b) spin superfluid mode.}
	\label{fig:PhaseVar}
\end{figure}


Figure \ref{fig:PhaseVar} shows the ground state when $a_{12}=90a_0$ for each component, shown in red and blue for the dipolar and nondipolar condensates, respectively. Overlaid, we have also plotted the phase variations for typical long-wavelength excitations on the density superfluid branch [Fig.~\ref{fig:PhaseVar} (a)] and the spin superfluid branch [Fig.~\ref{fig:PhaseVar} (b)], both with $k_zd/\pi\approx 0.10$. Figure \ref{fig:PhaseVar} (a) can be identified as a density-dominated superfluid mode by the strong phase gradient between domains, with the same signs (i.e. same superfluid velocity direction) in both components. Figure \ref{fig:PhaseVar} (b) can be distinguished by the fact that components have velocities in different directions. Interestingly, Fig.~\ref{fig:PhaseVar} (a) also exhibits backflow, identified by the negative gradient within the actual domains. The backflow velocity is weaker than the inter-domain gradient and so overall this mode maintains a superfluid character rather than a crystal one. Finally, we note that since the quantity $\bar{\mathcal{R}}_{\kappa}$ is a feature of the BdG excitation itself, its value does not change across BZs.

\subsection{Spin versus density periodicity}
\begin{figure*}
	\centering
	\includegraphics[width=0.8\textwidth]{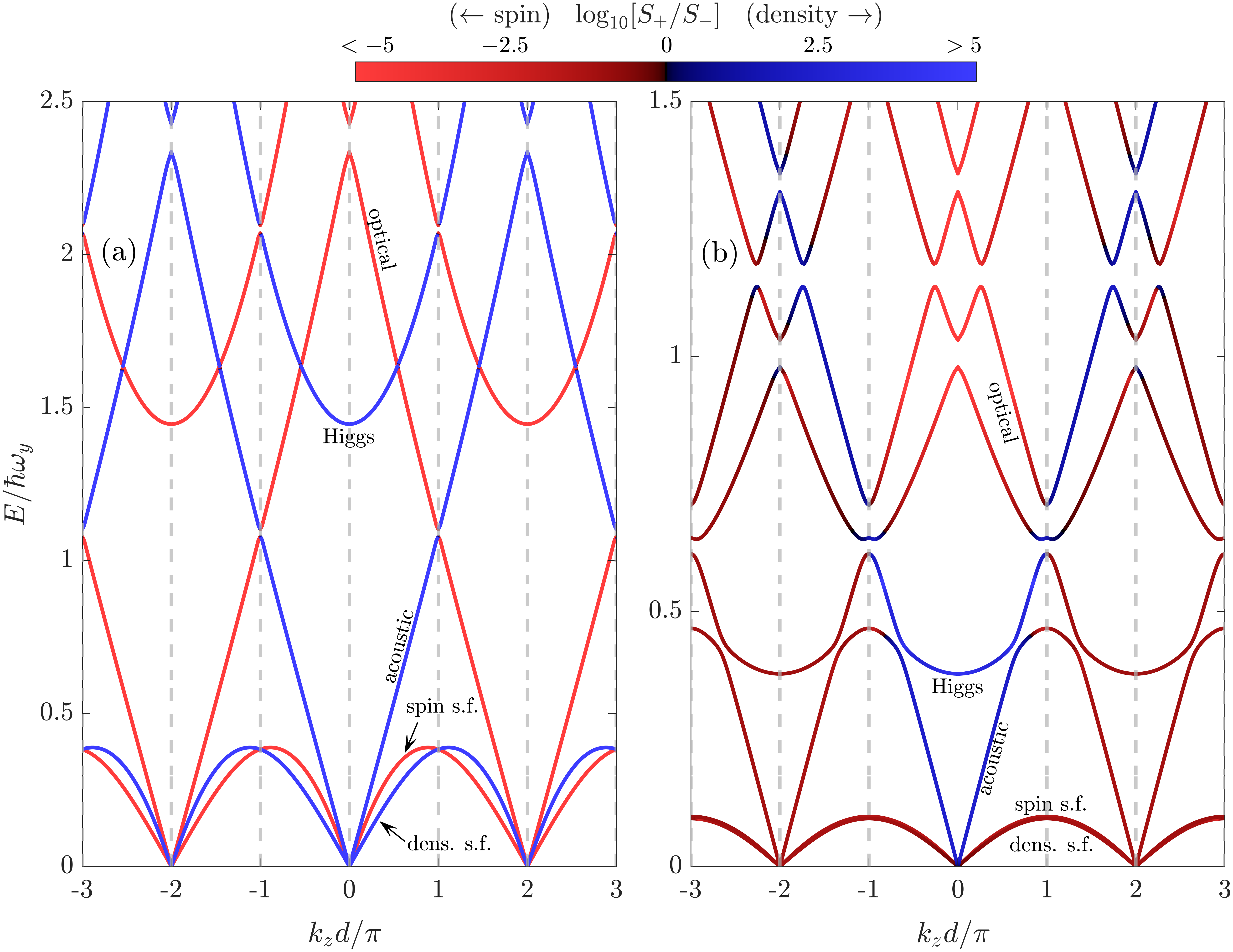}
	\caption{Density versus spin dynamic structure factor. (a) A balanced-component example with parameters: $(\mu_1^m,\mu_2^m) = (9.93,-9.93)\mu_{\rm B}$, $(a_{11},a_{12},a_{22})=(130,-25,130)a_0$, $(\bar{n}_1,\bar{n}_2) = (200,200) \mu{\rm m}^{-1}$. (b) The dipole-imbalanced case considered in Fig.~3 of the main text with parameters: $(\mu_1^m,\mu_2^m) = (9.93,0)\mu_{\rm B}$, $(a_{11},a_{12},a_{22})=(130,90,130)a_0$, $(\bar{n}_1,\bar{n}_2) = (200,200) \mu{\rm m}^{-1}$. }
	\label{fig:SFperiodicity}
\end{figure*}

For a given band, the flipping of spin versus density character across BZs can be understood by considering balanced systems. 
We define a balanced system to be one with equal average densities and intra-component interactions, i.e.~$a_{11}=a_{22}$ and $\mu_1^m=\pm\mu_2^m$. We consider a balanced system which becomes immiscible with a period of $L$. The BdG eigenvectors can be written in Bloch form as $u_{q,\nu,j}(z)=\bar{u}_{q,\nu,j}(z)e^{iqz}$, where we have separated the excitation label $\kappa$ into its band label $\nu$ and quasimomentum $q$, $\bar{u}$ has period $L$, with a similar equation for $v$. There is a translation symmetry between components, $\psi_2(z)=\hat{T}_{L/2}\psi_1(z)$, where $\hat{T}_{L/2}$ is translation operator of half unit cell length, and the eigenvectors obey the identity
$\bar{u}_{q,\nu,2}(z)=\pm \hat{T}_{L/2}\bar{u}_{q,\nu,1}(z)$, 
with $+$ for a density mode and $-$ for a spin mode, and similarly for $\bar{v}$.

To obtain dynamic structure factors (DSFs), we need to calculate the Fourier transforms of the spatial density ($+$) and spin-density ($-$) fluctuation terms
\begin{align}
    \delta n_{q,\nu,\pm} &= \psi_1(u_{q,\nu,1}-v_{q,\nu,1})\pm\psi_2(u_{q,\nu,2}-v_{q,\nu,2})\\
    &=e^{iqz}\Big(1+ s\hat{T}_{L/2}\Big)\psi_1(\bar{u}_{q,\nu,1}-\bar{v}_{q,\nu,1})
\end{align}
where $s=1$ for the spatial DSF of a spatial mode, or the spin DSF of a spin mode and $s=-1$ otherwise.
Writing $\psi_1(\bar{u}_{q,\nu,1}-\bar{v}_{q,\nu,1}) = \sum_n c_ne^{2\pi inz/L}$, $\delta n_{q,\nu,\pm} = \sum_n c_n[1+s(-1)^n]e^{i(2\pi n/L+q)z}$, i.e. only even $n$ contribute for $s=1$ and only odd $n$ contribute for $s=-1$, so the contribution to $S_+$ and $S_-$ alternates for each reciprocal lattice vector added.

As a demonstration of the periodically alternating spin versus density contributions, Fig.~\ref{fig:SFperiodicity} shows $S_+/S_-$ over several BZs for (a) a component-balanced case $(\mu_1^m,\mu_2^m) = (9.93,-9.93)\mu_{\rm B}$, and (b) the dipole-imbalanced case [$(\mu_1^m,\mu_2^m) = (9.93,0)\mu_{\rm B}$] considered in Fig.~3 of the main text. The contact interaction strengths have been altered in (a) so that both subplots are at a comparable distance from the unmodulated-to-supersolid phase transition and have similar superfluid fractions, with $f_s=0.45$ for (a) and $f_s=0.48$ for (b). The spin-density flipping is most striking for the balanced case [Fig.~\ref{fig:SFperiodicity} (a)], where a spin-density flipping occurs cleanly for every reciprocal lattice vector $k_zd/\pi = 2$ (here, $L=d$), i.e., every two BZs. Note also that spin and density branches cleanly cross one another without forming avoided crossings. The spin-density flipping partially survives for the imbalanced case [Fig.~\ref{fig:SFperiodicity} (b)], and coupling between spin and density branches is further evidenced by the presence of avoided crossings.

\end{document}